%% file: paper.tex
\documentclass{article}

\usepackage{geometry}
\usepackage[round,authoryear]{natbib}
\usepackage[hidelinks]{hyperref}
\usepackage{authblk}
\usepackage{tabularray}
\usepackage{graphicx}
\usepackage{amssymb}
\usepackage{xcolor}

\renewcommand\Affilfont{\small}

\makeatletter
\renewcommand\AB@affilsepx{;\hspace{0.6em}\protect\Affilfont}
\makeatother

\begin{document}

\title{Population-scale Ancestral Recombination Graphs with tskit 1.0}

\input{authors.tex}

\date{}
\maketitle


\noindent
Ancestral recombination graphs (ARGs) capture the full genetic history of
samples from a recombining species.
Although ARGs have been a central
theoretical object in population genetics for decades, their practical use was
constrained by the lack of scalable inference methods, standard
interchange formats, and software infrastructure. Recent breakthroughs in
simulation and inference have substantially changed this landscape,
leading to renewed interest in ARG-based analyses across population and statistical
genetics \citep{brandt2024promise,lewanski2024era,nielsen2024inference}.
The tskit library has played a key enabling role in this shift
and has become foundational infrastructure for working with ARGs.
This paper marks the release of tskit 1.0, which formalises long-term stability
guarantees for its data formats and APIs.

At the core of tskit is the succinct tree sequence data model which
defines a set of nodes (genomes at particular times) and
edges (inheritance relationships between nodes spanning genomic intervals)
in a simple tabular form \citep{kelleher2018efficient}.
This encoding provides a lossless representation of a general class of
ARGs, establishing a precise and machine-readable
definition suitable for large-scale computation \citep{wong2024general}.
The data model also incorporates site, mutation, population, and pedigree
information and supports arbitrary
metadata associated with each of these components of the ARG.
Provenance information is recorded natively, enhancing reproducibility
and transparency.
Together, these features make the tskit data model a
semantically complete and interoperable
representation of ARGs that serves as a common foundation across diverse
analytical workflows (Figure 1).


\begin{figure}
\includegraphics[width=\textwidth]{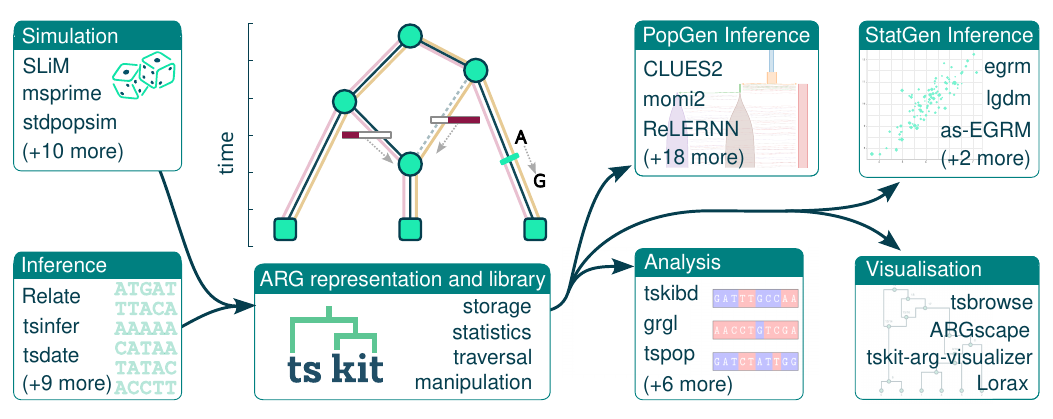}
\caption{Tskit enables an interoperable ARG software ecosystem.
ARGs produced by simulation or inference tools can be analysed by diverse
downstream applications via tskit’s well-defined tabular data model, C library and
Python/Rust/R bindings. Tools shown are representative examples from Table S1
(three per category; ordered by citation count).}
\end{figure}

Simulation is a fundamental tool in population genomics, and
was the first domain in which the tskit data model demonstrated its impact.
Introduced initially as part of the msprime simulator,
the tskit data model enabled performance improvements of
several orders of magnitude
over previous coalescent simulation approaches \citep{kelleher2016efficient}.
The same representation later
enabled efficient forward-time simulation of ARGs
and yielded substantial speedups by avoiding
explicit simulation of neutral mutations \citep{kelleher2018efficient}.
Because these forward-time and coalescent
simulators share a common underlying representation, their complementary
strengths can be combined within a single workflow. This has made it possible
to simulate ARGs under complex demographic scenarios involving geography and
selection that were previously infeasible, providing essential ground truth for
method evaluation. Simulation capabilities have continued to expand,
including whole-autosome ARG simulations for nearly 1.5 million individuals
based on a large human pedigree \citep{andersontrocme2023genes}.
A growing ecosystem of forward-time, coalescent, and hybrid simulation tools
builds directly on tskit (Table S1).


The lack of scalable inference methods has been a major obstacle
to practical application of ARGs. Although there
are many inference methods (see \citet{wong2024general}\ for a review),
tsinfer was the first to scale to hundreds of thousands of samples,
directly leveraging the tskit data model \citep{kelleher2019inferring}.
Many recent ARG inference methods have chosen to support tskit
as an output format in addition to their
own native representations (Table S1).
This shared output layer enables inferred ARGs to
interoperate directly with simulators, facilitating systematic evaluation and
benchmarking against known ground truth. It also shifts the burden of format
conversion away from downstream users, who can instead rely on inference tools
to emit results in a common, well-defined representation. The scalability and
flexibility of this approach are illustrated by the recent inference of an ARG
for 2.48 million SARS-CoV-2 whole genomes, which occupies 32 MiB of storage and
can be loaded into memory in under a second \citep{zhan2025pandemic}.


Efficient storage and analysis of large genetic datasets is a central
design goal of tskit, and the data model has enabled substantial
performance gains in downstream analyses.
For example, single-site
population genetic statistics can be computed orders of magnitude faster than from
genotype matrices while using far less memory by operating
on the underlying ARG structure \citep{ralph2020efficiently}.
Tskit exposes a large API with a
performance-critical core implemented in C and bindings available for Python,
Rust, and R.
Its vectorised, table-first design allows zero-copy access to
underlying arrays, supporting high-performance analysis pipelines.
As a result,
downstream tools inherit performance and correctness properties from a shared,
well-tested core.


The goal of tskit is to provide a shared technical foundation, centred on
efficient, well-tested, and thoroughly documented primitive operations on ARGs, rather
than to directly implement end-user workflows.
This design principle has enabled a broad ecosystem of downstream
software---spanning simulation, ARG inference, population and statistical genetic
inference, analysis, and visualisation---with 64 published tools now using tskit
as a core dependency (Table S1).
With the release of tskit 1.0, long-term stability of the data model
and APIs is explicitly guaranteed, reflecting a commitment to maintain this
minimal and composable core as the project evolves. By focusing on stable
primitives rather than prescribing analytical pipelines, tskit enables
methodological innovation to concentrate on modelling, inference, and
interpretation rather than bespoke data formats and tooling. In this way, tskit
1.0 provides a common and extensible foundation that supports the continued
expansion of ARG-based analyses as datasets, methods, and applications continue
to grow.
Extensive documentation, tutorials, and other information are available
at \url{https://tskit.dev}.

\subsection*{Acknowledgements}
We gratefully acknowledge funding from the Robertson Foundation,
the NIH (research grants HG011395 and HG012473),
and the NSF (research grant OAC-2104115),
supporting core tskit development.

\bibliography{paper}

\newpage

\section*{Supplementary Information}
\appendix
\setcounter{table}{0}
\setcounter{figure}{0}
\renewcommand{\thetable}{S\arabic{table}}
\renewcommand{\thefigure}{S\arabic{figure}}
\renewcommand{\thesubsection}{\arabic{subsection}}

This section contains the supplementary information, which we integrate
into the main document for ease of review.

\subsection*{Published tools using tskit}
We performed a search for published software using tskit.
To be included in this list, the software must be described in a publication
as a named tool intended for reuse and must use tskit.
This list is intentionally conservative, and does not include
(for example) publications depending on bespoke analysis
pipelines using tskit.
For each tool in Table S1, we list: its name (with a clickable
hyperlink to its source code repository);
the main implementation
language(s) of the tool based on GitHub summaries;
the DOI (with clickable hyperlink) of the publication in which this
tool was first described (or, in the case of SLiM, the first
publication in which tskit-based features were described);
and the year of publication and number of citations to this
publication according to data from OpenAlex
(\url{https://openalex.org}) retrieved on 2026-04-27.

\input{tools_table.tex}

\end{document}

%% file: authors.tex
\author[1,$\ast$]{ Ben Jeffery}
\author[1,$\ast$]{ Yan Wong}
\author[2,$\ast$]{ Kevin Thornton}
\author[3,4,$\ast$]{ Georgia Tsambos}
\author[1,$\dagger$]{ Gertjan Bisschop}
\author[5,$\dagger$]{ Yun Deng}
\author[6,$\dagger$]{ E. Castedo Ellerman}
\author[7,$\dagger$]{ Thomas B. Forest}
\author[8,$\dagger$]{ Halley Fritze}
\author[9,$\dagger$]{ Daniel Goldstein}
\author[10,$\dagger$]{ Gregor Gorjanc}
\author[11,$\dagger$]{ Graham Gower}
\author[12,$\dagger$]{ Simon Gravel}
\author[13,14,$\dagger$]{ Jeremy Guez}
\author[15,$\dagger$]{ Benjamin C. Haller}
\author[7,$\dagger$]{ Andrew D. Kern}
\author[16,$\dagger$]{ Lloyd Kirk}
\author[12,$\dagger$]{ Ivan Krukov}
\author[8,$\dagger$]{ Hanbin Lee}
\author[17,$\dagger$]{ Brieuc Lehmann}
\author[1,$\dagger$]{ Hossameldin Loay}
\author[18,$\dagger$]{ Matthew M. Osmond}
\author[19,20,21,$\dagger$]{ Duncan S. Palmer}
\author[7,$\dagger$]{ Nathaniel S. Pope}
\author[16,$\dagger$]{ Aaron P. Ragsdale}
\author[1,$\dagger$]{ Duncan Robertson}
\author[7,$\dagger$]{ Murillo F. Rodrigues}
\author[22,$\dagger$]{ Hugo van Kemenade}
\author[23,24,$\dagger$]{ Clemens L. Weiß}
\author[23,1,$\dagger$]{ Anthony Wilder Wohns}
\author[1,25,$\dagger$]{ Shing H. Zhan}
\author[19,$\dagger$]{ Brian C. Zhang}
\author[26]{ Marianne Aspbury}
\author[1]{ Nikolas A. Baya}
\author[7]{ Saurabh Belsare}
\author[27]{ Arjun Biddanda}
\author[28]{ Francisco Campuzano Jiménez}
\author[29]{ Ariella Gladstein}
\author[30,31]{ Bing Guo}
\author[1]{ Savita Karthikeyan}
\author[32]{ Warren W. Kretzschmar}
\author[33,34]{ Inés Rebollo}
\author[22]{ Kumar Saunack}
\author[35]{ Ruhollah Shemirani}
\author[36]{ Alexis Simon}
\author[7]{ Chris Smith}
\author[37]{ Jeet Sukumaran}
\author[8]{ Jonathan Terhorst}
\author[38]{ Per Unneberg}
\author[1]{ Ao Zhang}
\author[7,39,$\ddagger$]{ Peter Ralph}
\author[1,$\ddagger$]{ Jerome Kelleher}
\affil[1]{Big Data Institute, Li Ka Shing Centre for Health Information and Discovery, University of Oxford, OX3 7LF, UK}
\affil[2]{Department of Systems Biology, University of California, Irvine, CA 92697, USA}
\affil[3]{Melbourne Integrative Genomics, School of Mathematics and Statistics, University of Melbourne, Victoria, 3010, Australia}
\affil[4]{Department of Genome Sciences, University of Washington, Seattle, WA 98195, USA}
\affil[5]{Department of Genetics, Stanford University, Stanford, CA 94305, USA}
\affil[6]{Fresh Pond Research Institute, Cambridge, MA 02140, USA}
\affil[7]{Institute of Ecology and Evolution, University of Oregon, Eugene OR 97402, USA}
\affil[8]{Department of Statistics, University of Michigan, Ann Arbor, MI, 48109, USA}
\affil[9]{Khoury College of Computer Sciences, Northeastern University, MA 02115, USA}
\affil[10]{The Roslin Institute and Royal (Dick) School of Veterinary Studies, University of Edinburgh, EH25 9RG, UK}
\affil[11]{Microbiology and Infectious Diseases, SA Pathology, Adelaide, SA 5000, Australia}
\affil[12]{Department of Human Genetics, McGill University, Montreal, QC H3A 0C7, Canada}
\affil[13]{UMR 7206 Eco-Anthropologie, CNRS, MNHN, Université Paris Cité, 75116 Paris, France}
\affil[14]{Université Paris-Saclay, CNRS, INRIA, Laboratoire Interdisciplinaire des Sciences du Numérique, 91400, Orsay, France}
\affil[15]{Dept. of Computational Biology, Cornell University, Ithaca, NY 14853, USA}
\affil[16]{Department of Integrative Biology, University of Wisconsin--Madison, WI 53706, USA}
\affil[17]{Department of Statistical Science, University College London, London, WC1E 7HB, UK}
\affil[18]{Department of Ecology and Evolutionary Biology, University of Toronto, Toronto, Ontario, M5S 3B2, Canada}
\affil[19]{Department of Statistics, University of Oxford, OX1 3LB, UK}
\affil[20]{The Pioneer Centre for SMARTbiomed, Big Data Institute, Li Ka Shing Centre for Health Information and Discovery, University of Oxford, OX3 7LF, UK}
\affil[21]{Program in Medical and Population Genetics, Broad Institute of MIT and Harvard, Cambridge, Massachusetts 02142, USA}
\affil[22]{Independent researcher}
\affil[23]{Department of Genetics, Stanford University School of Medicine, Stanford, CA 94305, USA}
\affil[24]{Stanford Cancer Institute, Stanford School of Medicine, Stanford, CA 94305, USA}
\affil[25]{Infectious Disease Epidemiology Unit (IDEU), Nuffield Department of Population Health, University of Oxford, OX3 7LF, UK}
\affil[26]{Department of Paediatrics, University of Oxford, Oxford, UK}
\affil[27]{Department of Biology, Johns Hopkins University, Baltimore, MD, 21218, USA}
\affil[28]{Department of Biology, University of Antwerp, Antwerp, 2610, Belgium}
\affil[29]{Department of Human Genetics, University of California, Los Angeles, CA 90095, USA}
\affil[30]{Center for Vaccine Development and Global Health, University of Maryland School of Medicine, Baltimore, MD, 21201, USA}
\affil[31]{Institute for Genome Sciences, University of Maryland School of Medicine, Baltimore, MD, 21201, USA}
\affil[32]{Center for Hematology and Regenerative Medicine, Karolinska Institute, 141 83 Huddinge, Sweden}
\affil[33]{Instituto Nacional de Investigación Agropecuaria (INIA), Estación Experimental Las Brujas, Ruta 48 km 10, Canelones, Uruguay}
\affil[34]{Department of Statistics, Universidad de la República, College of Agriculture, Garzón 780, Montevideo, Uruguay}
\affil[35]{Institute for Genomic Health, Icahn School of Medicine at Mount Sinai, NY, 10029, USA}
\affil[36]{Sorbonne Université, CNRS, UMR 7144 AD2M, DiSEEM, Station Biologique de Roscoff, France}
\affil[37]{Biology Department, San Diego State University, San Diego, CA 92182-4614, USA}
\affil[38]{Department of Cell and Molecular Biology, National Bioinformatics Infrastructure Sweden, Science for Life Laboratory, Uppsala University, Husargatan 3, SE-752 37 Uppsala, Sweden}
\affil[39]{Department of Data Science, University of Oregon, Eugene OR 97402, USA}
\affil[$\ast$]{Joint first author}
\affil[$\dagger$]{Joint second author}
\affil[$\ddagger$]{Joint senior author}

%% file: tools_table.tex
\begin{tblr}[ 
        long,
        caption={Summary of 64 published software tools
            using tskit. See the supplementary
            text for details on the columns and inclusion criteria.}
     ]{ 
        colspec=lllrr,
        rowhead=1,
        width=\linewidth,
        colsep=3pt,
         }
\hline
Name & Language & Publication & Year & Cites \\
\hline
\SetCell[c=4]{c} \textbf{ Visualisation (4) }\\
\hline[dashed]
\href{ https://github.com/pratikkatte/lorax/ }{ Lorax }&
Python, JavaScript &
\href{ https://doi.org/10.64898/2026.02.19.706861 }{ 10.64898/2026.02.19.706861 }&
2026&
0\\
\href{ https://github.com/tskit-dev/tsbrowse }{ tsbrowse }&
Python &
\href{ https://doi.org/10.1093/bioinformatics/btaf393 }{ 10.1093/bioinformatics/btaf393 }&
2025&
1\\
\href{ https://github.com/chris-a-talbot/argscape }{ ARGscape }&
TypeScript, Python &
\href{ https://doi.org/10.48550/arXiv.2510.07255 }{ 10.48550/arxiv.2510.07255 }&
2025&
0\\
\href{ https://github.com/kitchensjn/tskit_arg_visualizer }{ tskit-arg-visualizer }&
Python, JavaScript &
\href{ https://doi.org/10.1093/bioadv/vbaf302 }{ 10.1093/bioadv/vbaf302 }&
2024&
0\\
\hline[dashed]
\SetCell[c=4]{c} \textbf{ Statistical Genetic Inference (5) }\\
\hline[dashed]
\href{ https://github.com/hanbin973/tslmm }{ tslmm }&
Python &
\href{ https://doi.org/10.1101/2025.07.14.664631 }{ 10.1101/2025.07.14.664631 }&
2025&
2\\
\href{ https://github.com/jitang-github/asegrm }{ as-eGRM }&
Python, C &
\href{ https://doi.org/10.1016/j.ajhg.2025.06.016 }{ 10.1016/j.ajhg.2025.06.016 }&
2025&
2\\
\href{ https://github.com/PalamaraLab/arg-lmm }{ arg-lmm }&
Python &
\href{ https://doi.org/10.1016/j.xgen.2025.101072 }{ 10.1016/j.xgen.2025.101072 }&
2025&
0\\
\href{ https://github.com/awohns/ldgm }{ lgdm }&
Python, Matlab &
\href{ https://doi.org/10.1038/s41588-023-01487-8 }{ 10.1038/s41588-023-01487-8 }&
2023&
39\\
\href{ https://github.com/Ephraim-usc/egrm }{ egrm }&
Python, C &
\href{ https://doi.org/10.1016/j.ajhg.2022.03.016 }{ 10.1016/j.ajhg.2022.03.016 }&
2022&
47\\
\hline[dashed]
\SetCell[c=4]{c} \textbf{ Analysis (9) }\\
\hline[dashed]
\href{ https://gitlab.gwdg.de/molsysevol/polarizing_without_outgroup }{ PolarBEAR }&
Python &
\href{ https://doi.org/10.1111/1755-0998.70105 }{ 10.1111/1755-0998.70105 }&
2026&
1\\
\href{ https://github.com/sgkit-dev/bio2zarr }{ bio2zarr }&
Python &
\href{ https://doi.org/10.1093/gigascience/giaf049 }{ 10.1093/gigascience/giaf049 }&
2025&
3\\
\href{ https://github.com/tskit-dev/tscompare }{ tscompare }&
Python &
\href{ https://doi.org/10.1093/genetics/iyaf198 }{ 10.1093/genetics/iyaf198 }&
2025&
2\\
\href{ https://github.com/HilaLifchitz/TwisstNTern }{ TwisstNTern }&
Python &
\href{ https://doi.org/10.32942/X29941 }{ 10.32942/x29941 }&
2025&
0\\
\href{ https://github.com/simonhmartin/twisst2 }{ twisst2 }&
Python &
\href{ https://doi.org/10.1093/genetics/iyaf181 }{ 10.1093/genetics/iyaf181 }&
2025&
0\\
\href{ https://github.com/bguo068/tskibd }{ tskibd }&
C++ &
\href{ https://doi.org/10.1038/s41467-024-46659-0 }{ 10.1038/s41467-024-46659-0 }&
2024&
33\\
\href{ https://github.com/aprilweilab/grgl }{ grgl }&
C++ &
\href{ https://doi.org/10.1038/s43588-024-00739-9 }{ 10.1038/s43588-024-00739-9 }&
2024&
11\\
\href{ https://github.com/lukashuebner/gfkit }{ gfkit }&
C++ &
\href{ https://doi.org/10.4230/LIPIcs.WABI.2024.5 }{ 10.4230/lipics.wabi.2024.5 }&
2024&
0\\
\href{ https://github.com/gtsambos/tspop }{ tspop }&
Python &
\href{ https://doi.org/10.1093/bioadv/vbad163 }{ 10.1093/bioadv/vbad163 }&
2023&
14\\
\hline[dashed]
\SetCell[c=4]{c} \textbf{ ARG Inference (12) }\\
\hline[dashed]
\href{ https://github.com/ps-pat/Moonshine.jl }{ Moonshine.jl }&
Julia &
\href{ https://doi.org/10.3389/fgene.2026.1753780 }{ 10.3389/fgene.2026.1753780 }&
2026&
0\\
\href{ https://github.com/popgenmethods/SINGER }{ SINGER }&
C++, Python &
\href{ https://doi.org/10.1038/s41588-025-02317-9 }{ 10.1038/s41588-025-02317-9 }&
2025&
14\\
\href{ https://github.com/YunDeng98/POLEGON }{ POLEGON }&
C++, Python &
\href{ https://doi.org/10.1073/pnas.2504461122 }{ 10.1073/pnas.2504461122 }&
2025&
1\\
\href{ https://github.com/simonhmartin/sticcs }{ sticcs }&
Python &
\href{ https://doi.org/10.1093/genetics/iyaf181 }{ 10.1093/genetics/iyaf181 }&
2025&
0\\
\href{ https://github.com/palamaralab/threads }{ Threads }&
C++, Python &
\href{ https://doi.org/10.1101/2024.08.31.610248 }{ 10.1101/2024.08.31.610248 }&
2024&
25\\
\href{ https://github.com/palamaralab/arg-needle }{ ARGneedle }&
Python, C++ &
\href{ https://doi.org/10.1038/s41588-023-01379-x }{ 10.1038/s41588-023-01379-x }&
2023&
115\\
\href{ https://github.com/davidrasm/Espalier }{ espalier }&
Python &
\href{ https://doi.org/10.1093/sysbio/syad040 }{ 10.1093/sysbio/syad040 }&
2023&
15\\
\href{ https://github.com/tskit-dev/sc2ts }{ sc2ts }&
Python &
\href{ https://doi.org/10.1101/2023.06.08.544212 }{ 10.1101/2023.06.08.544212 }&
2023&
14\\
\href{ https://github.com/tskit-dev/tsdate }{ tsdate }&
Python &
\href{ https://doi.org/10.1126/science.abi8264 }{ 10.1126/science.abi8264 }&
2022&
196\\
\href{ https://github.com/alimahmoudi29/arginfer }{ ARGinfer }&
Python &
\href{ https://doi.org/10.1371/journal.pcbi.1009960 }{ 10.1371/journal.pcbi.1009960 }&
2022&
37\\
\href{ https://github.com/MyersGroup/relate }{ Relate }&
C++ &
\href{ https://doi.org/10.1038/s41588-019-0484-x }{ 10.1038/s41588-019-0484-x }&
2019&
584\\
\href{ https://github.com/tskit-dev/tsinfer }{ tsinfer }&
C, Python &
\href{ https://doi.org/10.1038/s41588-019-0483-y }{ 10.1038/s41588-019-0483-y }&
2019&
379\\
\hline[dashed]
\SetCell[c=4]{c} \textbf{ Simulation (13) }\\
\hline[dashed]
\href{ https://github.com/tskit-dev/pyslim }{ pyslim }&
Python &
\href{ https://doi.org/10.1101/2025.09.30.679676 }{ 10.1101/2025.09.30.679676 }&
2025&
1\\
\href{ https://github.com/tskit-dev/tstrait }{ tstrait }&
Python &
\href{ https://doi.org/10.1093/bioinformatics/btae334 }{ 10.1093/bioinformatics/btae334 }&
2024&
10\\
\href{ https://github.com/bodkan/slendr/ }{ slendr }&
R &
\href{ https://doi.org/10.24072/pcjournal.354 }{ 10.24072/pcjournal.354 }&
2023&
15\\
\href{ https://github.com/samsonweiner/cnasim }{ cnasim }&
Python &
\href{ https://doi.org/10.1093/bioinformatics/btad434 }{ 10.1093/bioinformatics/btad434 }&
2023&
11\\
\href{ https://github.com/Trubenova/gridCoal }{ gridCoal }&
Python &
\href{ https://doi.org/10.1111/1755-0998.13676 }{ 10.1111/1755-0998.13676 }&
2022&
5\\
\href{ https://github.com/drewhart/geonomics }{ Geonomics }&
Python &
\href{ https://doi.org/10.1093/molbev/msab175 }{ 10.1093/molbev/msab175 }&
2021&
20\\
\href{ https://github.com/juntyr/necsim-rust }{ necsim-rust }&
Rust &
\href{ https://doi.org/10.48550/arXiv.2108.05815 }{ 10.48550/arxiv.2108.05815 }&
2021&
0\\
\href{ https://github.com/popsim-consortium/stdpopsim }{ stdpopsim }&
Python &
\href{ https://doi.org/10.7554/eLife.54967 }{ 10.7554/elife.54967 }&
2020&
248\\
\href{ https://catchenlab.life.illinois.edu/radinitio/ }{ RADinitio }&
Python &
\href{ https://doi.org/10.1111/1755-0998.13163 }{ 10.1111/1755-0998.13163 }&
2020&
56\\
\href{ https://github.com/eaton-lab/ipcoal }{ ipcoal }&
Python &
\href{ https://doi.org/10.1093/bioinformatics/btaa486 }{ 10.1093/bioinformatics/btaa486 }&
2020&
6\\
\href{ https://github.com/molpopgen/fwdpy11 }{ fwdpy11 }&
C++, Python &
\href{ https://doi.org/10.1534/genetics.119.302662 }{ 10.1534/genetics.119.302662 }&
2019&
80\\
\href{ https://github.com/MesserLab/SLiM }{ SLiM }&
C++, C &
\href{ https://doi.org/10.1093/molbev/msy228 }{ 10.1093/molbev/msy228 }&
2018&
882\\
\href{ https://github.com/tskit-dev/msprime }{ msprime }&
Python, C &
\href{ https://doi.org/10.1371/journal.pcbi.1004842 }{ 10.1371/journal.pcbi.1004842 }&
2016&
824\\
\hline[dashed]
\SetCell[c=4]{c} \textbf{ Population Genetic Inference (21) }\\
\hline[dashed]
\href{ https://github.com/blueraleigh/gaia }{ gaia }&
R, C &
\href{ https://doi.org/10.1126/science.adp4642 }{ 10.1126/science.adp4642 }&
2025&
12\\
\href{ https://github.com/Ephraim-usc/glike }{ gLike }&
Python, C &
\href{ https://doi.org/10.1038/s41588-025-02129-x }{ 10.1038/s41588-025-02129-x }&
2025&
10\\
\href{ https://github.com/a-ignatieva/dolores }{ dolores }&
Python &
\href{ https://doi.org/10.1093/molbev/msaf190 }{ 10.1093/molbev/msaf190 }&
2025&
6\\
\href{ https://github.com/osmond-lab/sparg }{ sparg }&
Python &
\href{ https://doi.org/10.1093/g3journal/jkaf214 }{ 10.1093/g3journal/jkaf214 }&
2025&
3\\
\href{ https://github.com/jthlab/phlash }{ phlash }&
Python &
\href{ https://doi.org/10.1038/s41588-025-02323-x }{ 10.1038/s41588-025-02323-x }&
2025&
2\\
\href{ https://github.com/a-ignatieva/spectre }{ spectre }&
Python &
\href{ https://doi.org/10.1093/genetics/iyaf184 }{ 10.1093/genetics/iyaf184 }&
2025&
1\\
\href{ https://github.com/aprilweilab/mrpast }{ mrpast }&
C++, Python &
\href{ https://doi.org/10.1101/2025.10.07.680347 }{ 10.1101/2025.10.07.680347 }&
2025&
0\\
\href{ https://github.com/osmond-lab/spacetrees }{ spacetrees }&
Python &
\href{ https://doi.org/10.7554/eLife.72177 }{ 10.7554/elife.72177 }&
2024&
22\\
\href{ https://github.com/kr-colab/mapNN }{ mapNN }&
Python &
\href{ https://doi.org/10.1111/1755-0998.14005 }{ 10.1111/1755-0998.14005 }&
2024&
4\\
\href{ https://github.com/LohseLab/gIMble }{ gIMble }&
Python &
\href{ https://doi.org/10.1371/journal.pgen.1010999 }{ 10.1371/journal.pgen.1010999 }&
2023&
36\\
\href{ https://github.com/nspope/coaldecoder }{ coaldecoder }&
C++, R &
\href{ https://doi.org/10.1073/pnas.2208116120 }{ 10.1073/pnas.2208116120 }&
2023&
31\\
\href{ https://github.com/kr-colab/disperseNN2 }{ disperseNN2 }&
Python &
\href{ https://doi.org/10.1186/s12859-023-05522-7 }{ 10.1186/s12859-023-05522-7 }&
2023&
12\\
\href{ https://github.com/racimolab/dinf }{ dinf }&
Python &
\href{ https://doi.org/10.1101/2023.04.27.538386 }{ 10.1101/2023.04.27.538386 }&
2023&
8\\
\href{ https://github.com/ctlab/gadma }{ GADMA }&
Python &
\href{ https://doi.org/10.1093/gigascience/giad059 }{ 10.1093/gigascience/giad059 }&
2022&
24\\
\href{ https://github.com/sunnyfangfangguo/SCAR_project_repo }{ SCAR }&
Python &
\href{ https://doi.org/10.1371/journal.pcbi.1010422 }{ 10.1371/journal.pcbi.1010422 }&
2022&
12\\
\href{ https://gitlab.com/mlgenetics/dnadna }{ dnadna }&
Python &
\href{ https://doi.org/10.1093/bioinformatics/btac765 }{ 10.1093/bioinformatics/btac765 }&
2022&
9\\
\href{ https://github.com/leospeidel/Colate }{ Colate }&
C++ &
\href{ https://doi.org/10.1093/molbev/msab174 }{ 10.1093/molbev/msab174 }&
2021&
72\\
\href{ https://github.com/simonharnqvist/DISMaL }{ DISMaL }&
Python &
\href{ https://doi.org/10.1016/j.tpb.2021.03.001 }{ 10.1016/j.tpb.2021.03.001 }&
2021&
7\\
\href{ https://github.com/kr-colab/ReLERNN }{ ReLERNN }&
Python &
\href{ https://doi.org/10.1093/molbev/msaa038 }{ 10.1093/molbev/msaa038 }&
2020&
182\\
\href{ https://github.com/avaughn271/CLUES2 }{ CLUES2 }&
Python &
\href{ https://doi.org/10.1371/journal.pgen.1008384 }{ 10.1371/journal.pgen.1008384 }&
2019&
179\\
\href{ https://github.com/popgenmethods/momi2 }{ momi2 }&
Python &
\href{ https://doi.org/10.1080/01621459.2019.1635482 }{ 10.1080/01621459.2019.1635482 }&
2019&
179\\
\hline
\end{tblr}